\documentstyle[11pt,newpasp,twoside,epsfig]{article}
\def\edcomment#1{\iffalse\marginpar{\raggedright\sl#1\/}\else\relax\fi}
\reversemarginpar

\begin{document}
\newcommand{\sco}{Scorpius~X-1}
\newcommand{\her}{Hercules~X-1}
\newcommand{\hzher}{HZ Herculis}
\newcommand{\cyg}{Cygnus~X-2}
\newcommand{\xte} {\textit{RXTE}}
\newcommand{\rxte}{\textit{RXTE}}

\title{Rapid optical spectroscopy of X-ray binaries with Keck II}
\author{Kieran O'Brien}
\affil{Astronomical Institute ``Anton Pannekoek'', University of Amsterdam, 1098-SJ Amsterdam, The Netherlands}

\begin{abstract}
iThe large collecting area of 4+ metre class telescopes allows atronomers
to observe bright objects on faster timescales than previously possible.
In order to exploit this capability, we designed a novel read-out mode for
use on the 10-m Keck-II telescope on Mauna Kea, Hawaii. This took the form
of a continuous read-out mode, enabling us to obtain a target spectrum
between 3500 and 8500 \AA, with a dispersion of ~2.5 \AA/pix, every *70
milliseconds*.

I present results from 5 nights of observations of Interacting
Binaries, using this novel data acquisition system in July 1998. During
the run we obtained more than 1.2 million optical spectra, many of which
were simultaneous with pointed RXTE observations. This fast spectroscopic
capability has allowed us to explore regimes previously only accessible
with broadband instruments. I will concentrate on results from the
coordinated Keck/RXTE observations, which have revealed interesting
correlated variability on timescales from seconds to days. I will also show
the results of a spectroscopic-timing analysis of QPOs in Hercules X-1,
which shows evidence for the decay of individual blobs of material
orbiting the neutron star.
\end{abstract}

\section{Introduction}

In the last 5+ yrs, with the increasing availability of 4+ metre optical telescopes and fast CCDs it has become possible to perform high time resolution optical spectroscopy. Fast read-out modes are becoming available on these telescopes allowing 1-2k pixel spectra to be read out on timescales of seconds or faster. This allows us to use spectroscopy to explore the rapid optical variability of relatively bright objects (V $\la$ 15), for instance interacting binaries, where previously only broad-band photometry had been possible.

There are currently two ways to perform such observations. The first uses a drift mode, whereby the exposed region of the CCD is rapidly shifted to an available space on the CCD near the serial register of the CCD. The number of spectra that can be taken in one exposure in this mode is limited by either the physical size of the CCD, or the size of the memory buffer. Once this limit has been reached the data is read-out and a new exposure can begin. This read-out can takes $\sim$ 10 seconds and leads to poor duty cycles for the fastest observations (for example 2048$\times$19.2 millisecond observations, which is standard for the VLT takes 39.35 seconds, but is followed by a read-out time of 92 seconds.) An alternative method uses a continuous-clocking mode, where the CCD is read-out constantly. Unlike the drift mode, there is no limit to the number of spectra that can be readout ina  given exposure and deadtime is kept to a minimum. Currently only drift modes are available for visiting observers, examples of these types of modes are

\begin{itemize}

\item High Time resolution (HIT) mode using the FOcal Reducer Spectrograph (FORS2) on the Very Large Telescope at Cerro Paranel in Chile. This drift mode allows up to 2048 spectra to be shifted down the CCD before read-out. A number of fixed exposure times are available, between 1 and 300-msec. (A continuous clocking mode will be available soon.)

\item Low Smear Drift mode using ISIS on the William Herschel Telescope on La Palma. This drift mode allows is limited by the size of the memory on the DMS (16Mb). It allows read-out of several spatial pixels in a given exposure. 

\item Drift mode using RGO spectrograph on the Anglo-Australian Telescope at Siding Spring, New South Wales. This mode is similar to the LSD mode on the WHT, again using a physical buffer which limits the number of spectra that can be obtained in one exposure.

\end{itemize}

A continuous clocking mode works in a similar manner to drift mode systems, however it doesn't suffer from time-consuming read-out the data, as is the case with existing systems. This is critical in reducing the overheads of such systems, especially when working at the fastest rates, as highlighted earlier. The continuous clocking system used in our project was developed by Richard Gomer for use with the Low-Resolution Imaging Spectrograph (LRIS; {Oke, 1995}) at the Cassegrain focus of the 10 metre Keck II telescope, situated near the summit of Mauna Kea, Hawaii. In the next sections, I will describe the data we obtained using this system, including details of the data reduction necessary and highlight some of the results obtained.

\section{The project}
The three main goals of the project were

\begin{itemize}

\item Echo mapping of X-ray binaries

\item Correlated long timescale variability ($\sim$ hours)

\item Spectroscopic timing studies

\end{itemize}

Echo mapping is an indirect imaging technique that uses the time delays between X-ray and optical variability to map the reprocessing regions in an X-ray binary (see O'Brien and Horne (2001a) for a review of echo-mapping of X-ray binaries). The optical emission arises from the reprocessing of X-ray irradiation and the time of flight difference between the directly observed X-rays and the subsequent reprocessed optical emission constrains the position of the reprocessing region. This technique has been used to map the broad line region in Active Galactic Nuclei (Horne 2001) and more recently adapted to X-ray binaries to constrain the positions of the reprocessing sites in such systems (Hynes et al. 1998; O'Brien et al. 2001d). However, such projects have been limited to short duration photometric observations, whilst \xte\ and EXOSAT have allowed us to accurately observe the X-ray variability on timescales of milliseconds. Our goal is to search for time delayed echoes of the rapid X-ray variability and use this to map the accretion flow in X-ray binaries.

Perhaps the most exciting consequence of using optical spectroscopy for echo-mapping is the ability to create physical parameter echo-maps (eg. temperature and density maps) by comparing the time delays from different emission lines. Such maps could be used to constrain models for the accretion flow, for instance helping to determine the role played by Advection Dominated Accretion Flows (ADAFs) on the inner regions of the accretion flow.

Correlated long timescale variability is vital in understanding the multi wavelength picture of X-ray binaries, in order to improve phenomenological models of such systems. Photometry shows us that rapid variability occurs on many/all timescales and traditional spectroscopy reveals a number of different emission lines in the optical spectra. Spectra of sufficient time resolution will allow us to understand the rapid response of such systems and whether or not this is driven by the X-ray irradiation. 

Spectroscopic timing studies will not only reveal the existence of any periodic or quasi-periodic variability in such systems, but also to constrain the production mechanism and site using the spectrum of the variability. In the case of X-ray binaries this will allow us to search for analogues of the QPOs seen in the X-ray power spectra of such systems. 

\section{The data acquisition system}


The data were obtained during 5 clear nights UT 02-06 July 1998. During the observations over 1.2 million spectra were obtained. The principal targets were \cyg, \her\ and \sco, for which the observations were simultaneous with \rxte\ to search for correlated X-ray and optical variability. Additional targets were observed when X-ray observations were not possible (earth occultation, SAA passage). Table~1 is a summary of the observations, including the approximate number of optical spectra taken on each target and a brief description of each object.

\begin{table}
\begin{center}
\begin{tabular}{ccccc}
\tableline
Object & Night & Number of & \rxte & Object type \\
name & observed & optical spectra & coverage? & \\ \tableline
\cyg & 1,2,4,5 & 504,000 & Y & LMXB, Z-source \\
\her & 2,5 & 105,000 & Y & Binary X-ray Pulsar \\
\sco & 1,2,3,4 & 355,000 & Y & LMXB, Z-source \\
AE Aqr & 2,4,5 & 87,000 & N & Magnetic CV \\
V2051 Oph & 3,4 & 89,000 & N & Dwarf Nova \\
SS Cyg & 2,5 & 50,000 & N & Dwarf Nova \\
AM Her & 2 & 25,000 & N & Polar \\
V404 Cyg & 5 & 14,000 & N & Black Hole Binary \\ \tableline \tableline
\end{tabular}
\end{center}
\caption[Summary of all targets for Keck observations]{Summary of the target list of the observations.}
\end{table}

\subsection{Technical description} 

The data system works by creating an invisible tap on the fibre optic cable between the VME and timing boards, to intercept and record the data from the CCD in the form of a continuous byte stream. With the data system switched on, the CCD is set into continuous read-out mode, where the rows of the CCD are shifted down the CCD and readout every $\sim$ 0.072 seconds, fixing the time resolution of the data system. The data can be thought of as coming from a virtual CCD with 2148 columns (including 25 pixels of under-scan and 75 pixels of over-scan) and $(A + T/0.072)$ rows, where $T$ is the duration of the run in seconds and $A$ is the row number, from the serial read-out, where the image of the slit falls on the CCD. The spatial dispersion is 4.6 pixels/arcsecond, so with on-chip binning of 8, each row covers 1.74 arcseconds. A 5.2 arcsecond slit was used to ensure that all the incident starlight was collected. The slit was partially covered using aluminized mylar tape to create a 5.2x5.2 arcsecond square aperture. Thus the image of the aperture covers three rows of pixels on the CCD, one row containing the object and two rows containing sky. Since we have no information as to the exact location of the image on the CCD our uncertainty in $A$ is $\pm 1$, leading to an intrinsic uncertainty in the absolute timing of $\pm 0.072$ seconds. It also leads to a correlation between adjacent spectra, since three pixels are being exposed simultaneously, however this is low since the seeing was $\sim 0.8$ arcseconds during the observations.

Two time reference systems were used. First, an internal clock which updated roughly every 55 ms was used to mark the time on each 2148-pixel spectrum. This clock was used to display the time on the monitor, and the displayed time was compared to the audible WWVH time signal received by a short wave radio. The offset could be determined to roughly 0.1 seconds, and this was recorded roughly every 30 minutes. 

Due to the nature of the continuous byte stream of data, it was impossible to obtain accurate times for each individual spectrum. In order to find the times accurately timestamps were placed at several points throughout the observations. These timestamps were created using an incandescent lamp that illuminated the CCD at a known time, from which we found an ephemeris and exposure time. Individual time marks, accurate to approximately 200ms, were placed after every other spectrum using the computer clock. These were used as a check for our calculated ephemeris and exposure time. 

\subsection{Sky subtraction}
Sky spectra were taken at the beginning and end of each run, by moving the telescope, so that long timescale variations could be detected. The mean and variable components of the sky spectrum were found by creating a lightcurve for each pixel and extracting the mean (coefficient 1) and gradient (coefficient 2) of this lightcurve. The average gradient was found to be $< 5 \times 10^{-7}$ counts/spectrum. Each run consisted of $\sim 40,000 - 80,000$ spectra, so that the expected change in the sky level during the run is $\sim$ 0.1\%. The gradient was therefore set to zero during our analysis so that no spurious pixel to pixel variations were introduced. The mean value was calculated for each run and filtered in wavelength, with a running median filter of width 11 pixels. 

This mean sky spectrum also shows several electronic features. The origin of these features is unknown, but they appear to be removed very well from the object spectra if modeled as sky features. During the sky subtraction, the variance of these points was increased so that their weighting in subsequent analysis was reduced.

\subsection{Wavelength calibration}
The arc calibration was done by fitting a second order polynomial to 7 identified lines in a median spectrum of 1400 individual exposures of Hg and Ar lamps. The individual arc lines are broad due to the large slit width used in our observations (5.2 arcseconds, or 60 $\AA$), however the Gaussian fits to the lines find the centroid well (This can be seen by comparing the calculated wavelengths of spectral lines to their known wavelengths). Arc spectra from the beginning and end of the exposures were used to take into account any drifts in the wavelength scale. The wavelength calibration was applied using the \textsc{molly} spectral analysis package.

\subsection{Flux calibration}
The individual spectra on a given night were flux calibrated using exposures taken of the standard star, Feige 67 (Oke 1990). A low order polynomial fit was found to the median of 2000 individual spectra and this calibration was applied using the subroutine \textsc{fcal} in the analysis package \textsc{molly}. Unfortunately, due to the sudden drop off in the response of the grating used, it was not possible to fit to the standard star below $\sim$ 3600\AA, thus reducing our effective wavelength coverage. 

\section{Summary of results}

While the thorough analysis of such a vast and unique dataset could take many years, I wish to briefly summarize some of the initial findings of our work. These results highlight progress made towards each of the three goals of the project.

\subsection{Correlated X-ray and optical variability in Cygnus X-2}

We have created simultaneous X-ray and optical lightcurves from our dataset, including continuum and line lightcurves from the optical spectra in order to investigate the correlations in the system. We have found correlations on the timescales of hours that are shown in Figures 1 and 2, however these correlations are not as simple as one would expect. In the simple model of reprocessing, hard X-ray penetrate below the optical photosphere, heating these regions, which subsequently emit optical and UV photons as they cool. The soft X-ray photons are absorbed above the optical photosphere and result in emission line photons. Both of these processes should lead to a direct relation between the X-ray and optical fluxes, once the recombination and cooling timescales are taken into account. However, we have observed that while this direct relation holds when the source is on the Normal Branch (hard X-ray spectrum, medium intensity), it does not hold on the Flaring branch (soft X-ray spectrum, low intensity), as illustrated in Figure 1. However the correlated variability appears stronger on the flaring branch, where the X-ray variability is greater, which appears to contradict the previous observation of an anti-correlation in this state. Further work is needed in order to understand this phenomenon, however it appears that the X-ray flux is comprised of more than one component, one of which is responsible for the short timescale ($\sim$ 10's seconds) correlations and another of which is responsible for the longer variability ($\sim$ minutes-hours) which has no rapid response in the optical. The origin of these components remains unclear, although it is possibly related to the balance between the accretion and outflow rates (O'Brien 2001c), which varies as the state changes in the system and are expected to have different spectra. 

\begin{figure}
\plotfiddle{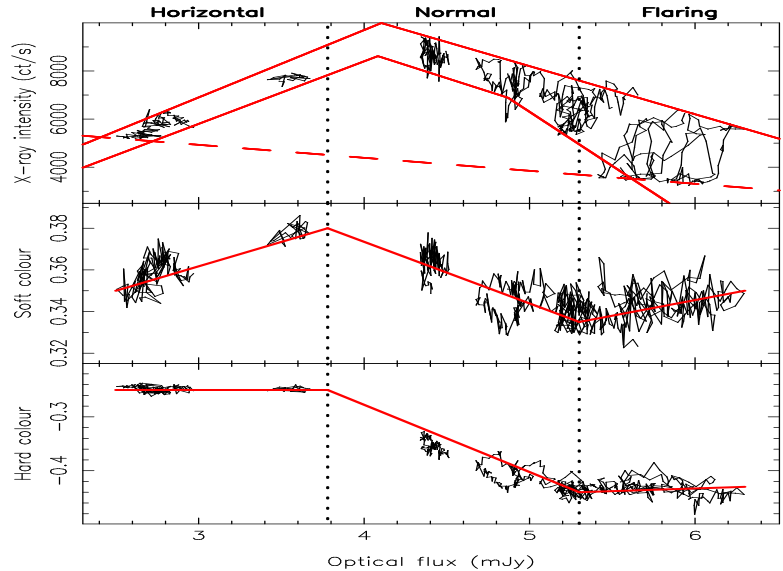}{2.7in}{0}{55}{30}{-170}{-15}
\caption{The correlated variability of the optical continuum flux (5000-5800 $\AA$) and the X-ray intensity, Soft colour and Hard colour. The Horizontal, Normal and Flaring regions refer to the branches on the Z-curve, determined through fits to the Z-shape in the X-ray colour-colour diagram.}
\end{figure}

\begin{figure}
\plotfiddle{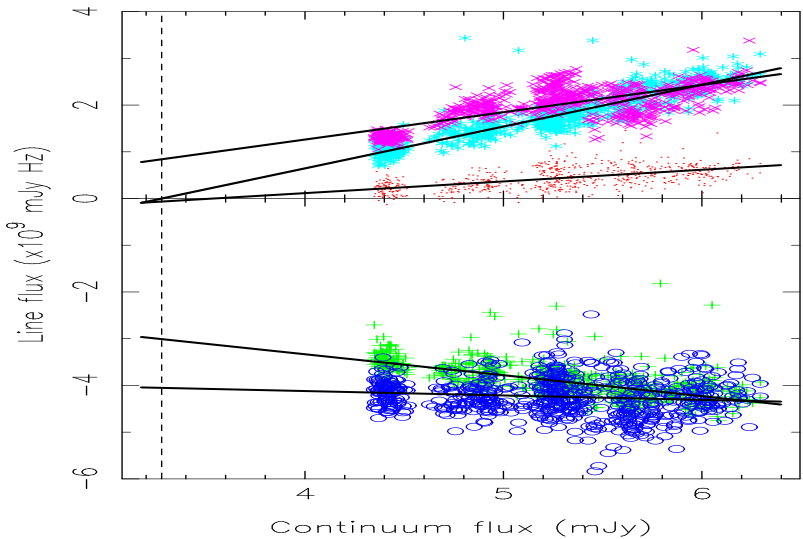}{2.5in}{0}{65}{30}{-185}{-15}
\caption{A plot showing the correlation between the continuum flux and various lines present in the optical spectrum of the source. The lines shown are H$\,\alpha$ (dots), H-$\,\beta$ (pluses), H-$\,\gamma$ (ellipses), Bowen blend (stars) and HeII (crosses). The thick lines are best fit to the individual lines. The dashed line represents zero irradiation, when the Bowen blend line flux is zero.}
\end{figure}

\subsection{Echo-mapping of Cygnus X-2}

In order to perform echo-mapping of Cygnus X-2 it will be necessary to decompose the X-ray variability into the individual components described in the previous section. By removing the slowly varying component we will determine the lightcurve of the variability that drives the rapid response. The correlations can be seen in teh non-deconvolved lightcurve shown in Figure 3 during the first set of observations around t=77000 secs. Preliminary analysis of these lightcurves reveal a time delay of $\sim$ 60 secs in the cross-correlation functions. This is in the delay expected if the reprocessing is taking place in the outer regions of the accretion disc. The peak will come from reprocessing at the rear of the disc, where the projected area of the reprocessing region is greatest (O'Brien 2000). This supports the work of Casares et al. (1998), who found little evidence for irradiation on the innerface of the companion star. 

\begin{figure}
\plotfiddle{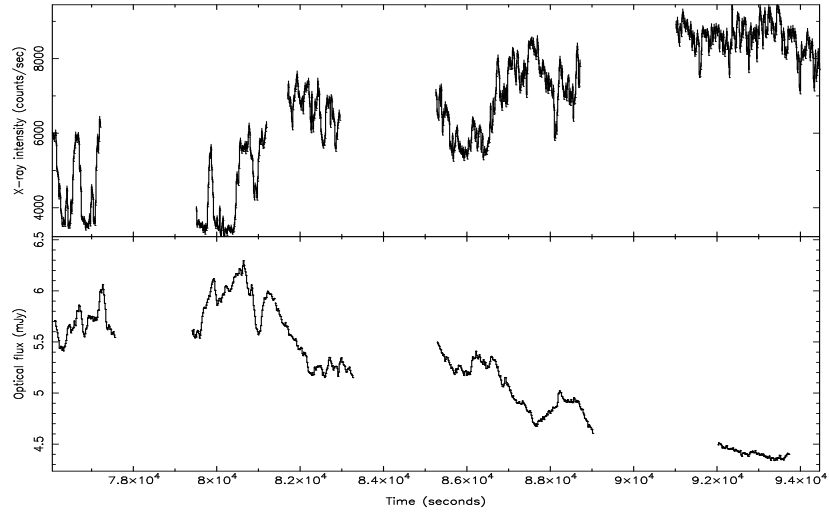}{2.2in}{270}{45}{35}{-180}{195}
\caption{The simultaneous X-ray (top) and optical continuum (bottom) lightcurves for Cygnus X-2, during the fifth night of our observations. The time is seconds after JD 2451000.}
\end{figure}

\subsection{Optical mHz QPOs in Hercules X-1}

We observed a low coherence QPO in the continuum and lines of Hercules X-1 (O'Brien et al. 2001b). The power spectrum was fitted with 3 components (White noise, power law and a QPO) and the central frequency of the QPO determined to be 35mHz (see Figure 4). The dynamic power spectrum, shown in Figure 5, created using the Gabor transform method (Boyd ????), shows the QPO is comprised of several higher coherence QPOs with a range of frequencies. The coherence of these individual features are higher and are associated with a dynamical timescale in the disc. The central frequency is similar to that observed by Boroson et al. (2000) in the UV lightcurve of Her X-1 and is tentatively associated with the impact point of the accretion stream, incident on the twisted, precessing accretion disc. No evidence was found for this QPO in the simultaneous X-ray power spectrum, although the observations took place during the X-ray ``off-state'' of the 35-day super-orbital periodicity.

\begin{figure}
\plotfiddle{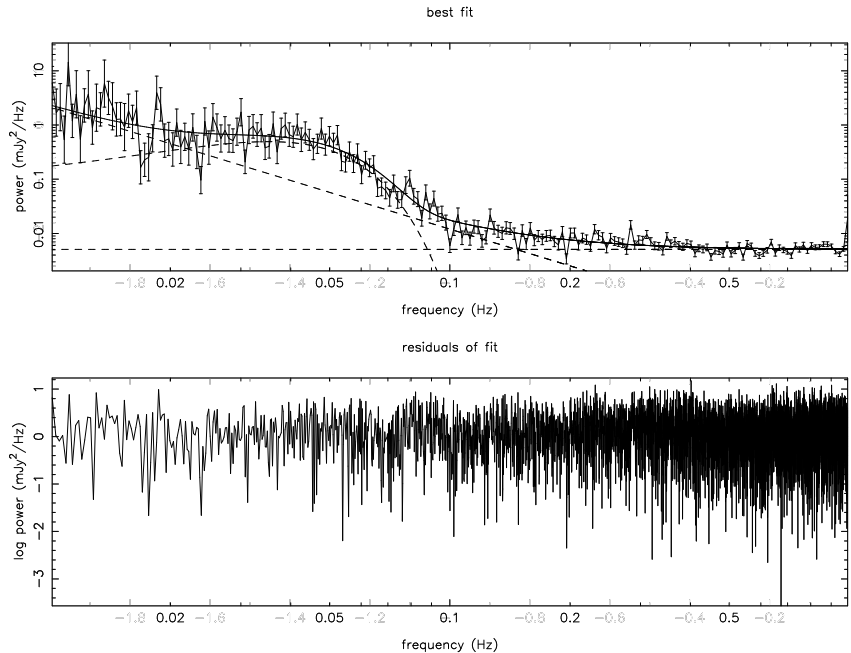}{3.1in}{270}{45}{45}{-170}{255}
\caption{The binned power spectrum of the optical continuum flux of Hercules X-1. The dashed lines represent the best fit white noise, power law and QPO components. The solid line shows the sum of these three components. The bottom panel shows the residuals to the fit, with the full sampling of the power spectrum.}
\end{figure}

\begin{figure}
\plotfiddle{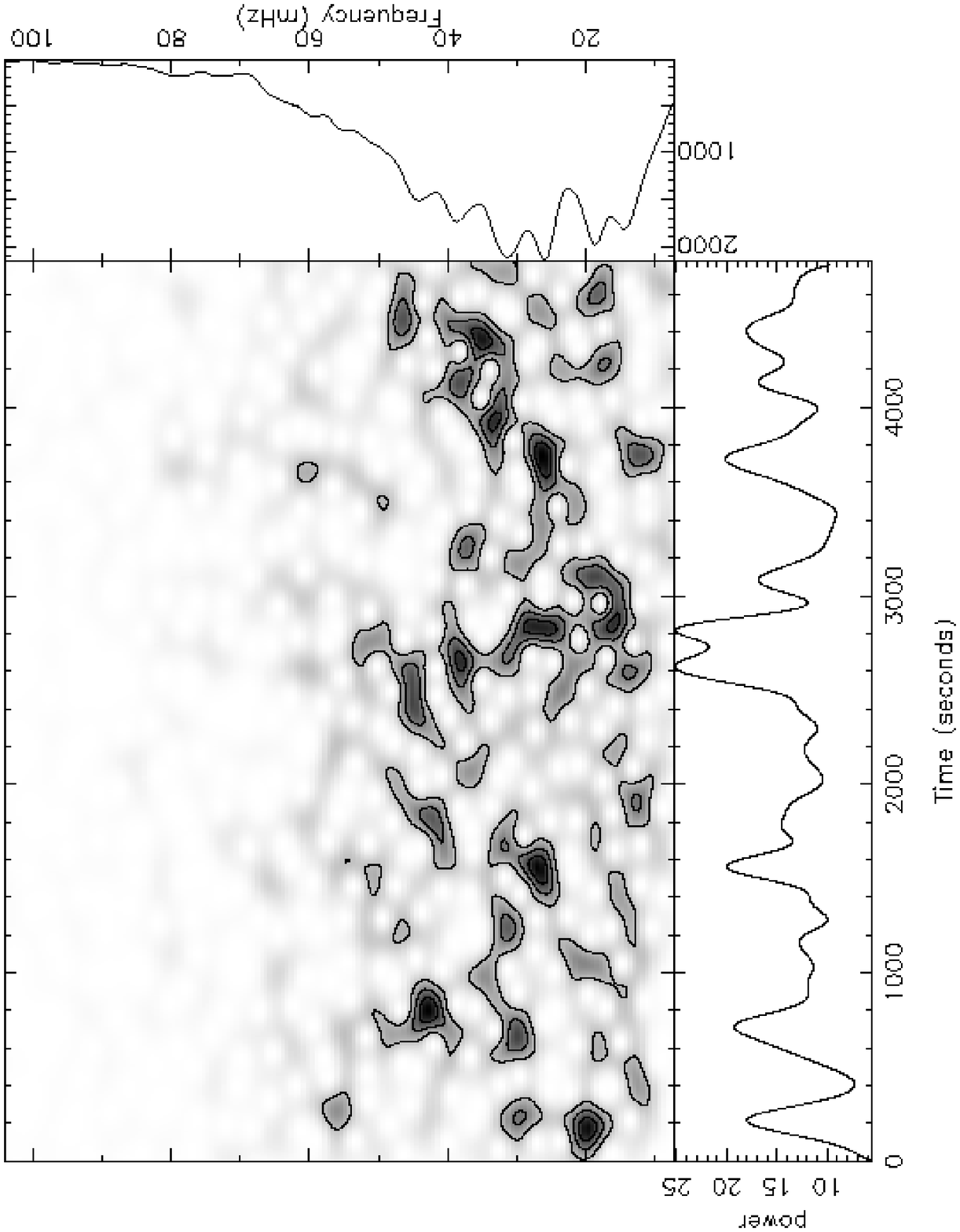}{3.1in}{270}{45}{45}{-170}{255}
\caption{The dynamic power spectrum of the continuum variations in Hercules X-1. Individual short-lived oscillations, centred around the QPO frequency can be seen. The lower panel shows power as a function of time and the right-hand panel shows the average power spectrum.}
\end{figure}

\subsection{Emission line oscillations in V2051 Ophiuchi}

During a short observation, covering only 200seconds, Steeghs et al. (2000) observed continuum and line emission oscillations in the eclipsing dwarf nova V2051 Ophiuchi. These observations took place during the decline from a normal outburst of this system and included observations during an eclipse, enabling a determination of the spectrum of the white dwarf, which is difficult to decompose with slower spectroscopic observations. 

The continuum oscillations with a period of 56.12 seconds are associated with the spinning white dwarf, while emission line oscillations with a period of 29.77 seconds are associated with a region $\sim$ 12 white dwarf radii from the white dwarf. The form of the emission line oscillation is shown in Figure 6, which shows a clear blue-red shift pattern. Steeghs et al. (2000) identify this with the vertical structure in the disc at the circularisation radius, where the accretion stream impacts onto the disc. 

\begin{figure}
\plotfiddle{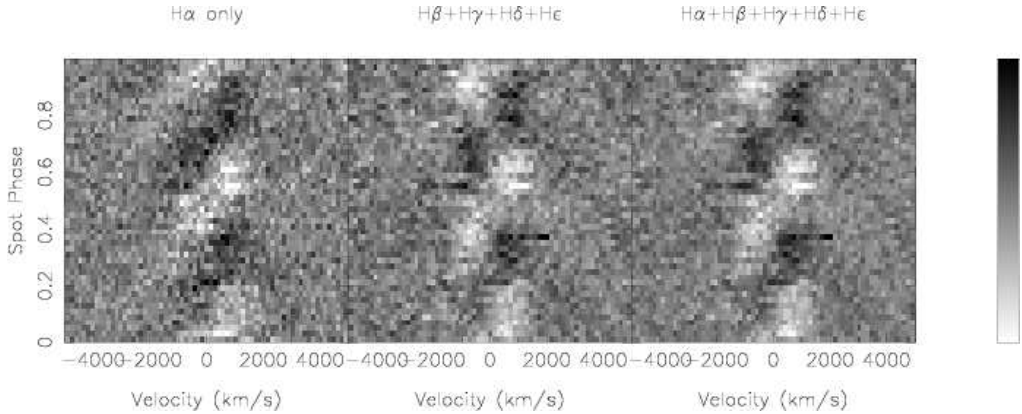}{1.9in}{270}{55}{55}{-220}{240}
\caption{Data folded on the oscillation period of 59.53 secs. Left, H-alpha only, middle and left, several Balmer lines have been combined to increase the signal. (Figure 8. of Steeghs et al. 2001)}
\end{figure}

\section{Summary and future}
We have obtained low resolution spectra of interacting binaries with integration times of milliseconds and little or no deadtime with a novel continuous clocking data acquisition mode and the current hardware available on the Keck II telescope. These observations have revealed a number of interesting phenomena, including time-delayed optical variability that is being driven by X-ray irradiation in the LMXB Cygnus X-2. We have studied the long timescale correlations between the X-ray and optical spectra in Cygnus X-2 as it changes X-ray state, as defined by its position in the X-ray colour-colour diagram. We have also studied QPOs in the continuum and lines of the X-ray pulsar Hercules X-1 and the dwarf nova V2051 Ophiuchi.

These observations highlight some of the interesting phenomena that can be explored with rapid spectroscopy of systems like X-ray Binaries and Cataclysmic Variables and how such observations can help deepen our knowledge of existing phenomena and explore a range of new phenomena in a relatively unexplored region of parameter space. However, equally importantly, they show how such studies can be performed with existing hardware during bright time. This increases the productivity of large telescopes during these periods without the need for expensive, dediated instruments. This dataset shows that large telescopes have the under-used ability to allow astronomers to go deeper {\bf and} faster, a capability that should not be overlooked.

\acknowledgments

The author would like to thank the many collaborators in this project; especially Keith Horne, Richard Gomer, Bev Oke, Danny Steeghs and Warren Skidmore who have all contributed directly to this work. We also thank John Cromer for writing, testing, and loading the software that allowed the LRIS CCD to read out continuously, and Bob Leach for helpful discussions.  We especially thank Tom Bida and Frederic Chaffee for kindly letting us make changes to the LRIS system.

\end{document}